\documentstyle[emulateapj]{article}

\begin{document}
\newcommand{\kms}{$\rm {km}~\rm s^{-1}$}
\newcommand{\Ho}{H$_\circ$}
\newcommand{\qo}{q$_\circ$}

\slugcomment{Submitted to the {\it Astrophysical Journal Letters}}

%\slugcomment{ Draft Version 2.0, \today}
%\pagestyle{myheadings}
%\mark{DRAFT version 2.0: \today}

\title{Wide Field Imaging of the Hubble Deep Field-South Region I:
Quasar Candidates
\footnote{Based on observations obtained at Cerro Tololo
Inter-American Observatory, a division of the National Optical
Astronomy Observatories, which is operated by the Association of
Universities for Research in Astronomy, Inc.\ under cooperative
agreement with the National Science Foundation.}
}

\author{Povilas Palunas\altaffilmark{1}$^,$\altaffilmark{2},
Nicholas R. Collins\altaffilmark{3},
Jonathan P. Gardner,
Robert S. Hill\altaffilmark{3},
Eliot M. Malumuth\altaffilmark{3}, 
Alain Smette\altaffilmark{4},
Harry I. Teplitz\altaffilmark{4},
Gerard M. Williger\altaffilmark{4},
Bruce E. Woodgate}
\affil{Laboratory for Astronomy \& Solar Physics, Code 681,
NASA/Goddard Space Flight Center, Greenbelt, MD 20771\\
palunas@gsfc.nasa.gov,
collins@zolo.gsfc.nasa.gov,
gardner@harmony.gsfc.nasa.gov,
bhill@virgil.gsfc.nasa.gov,
eliot@barada.gsfc.nasa.gov,
asmette@band3.gsfc.nasa.gov,
hit@binary.gsfc.nasa.gov,
williger@tejut.gsfc.nasa.gov,
woodgate@s2.gsfc.nasa.gov
}

\altaffiltext{1}{NASA/NRC Resident Research Associate}

\altaffiltext{2}{Department of Physics, Catholic University of America, Washington, DC 20064.}

\altaffiltext{3}{Raytheon ITSS, 4400 Forbes Blvd., Lanham MD 20706}

\altaffiltext{4}{NOAO Research Associate}

\begin{abstract}
We present candidate quasars from a multi-color ($uBVRI$ +
narrow-band) imaging survey of 1/2 square degree around the Hubble
Deep Field -- South. We identify 154 candidate quasars with $B < 23$
using color selection, consistent with previously measured QSO number
counts if we assume a 60\% selection efficiency.  The narrow-band
filter (NB) was centered at 3958\AA\ to detect Ly$\alpha$ at the
redshift of J2233--6033, the HDF--S QSO. We confirm the presence of
Ly${\alpha}$ nebulosity extending $\sim12\arcsec$ around the HDF--S
QSO, reported by Bergeron et~al. (1999). We detect 10 point-like
objects in emission through the NB filter. Of these, 7 satisfy our QSO
color selection criteria.  One of emission-line objects is a
$B\sim20$ radio-quiet quasar at $z=1.56$, 6.7\arcmin\ from the line of
sight to the HDF--S QSO and $\sim$12\arcsec\ from the western edge of
the WFPC2 deep field.

\end{abstract}


\section{Introduction}

The recent Hubble Deep Field-South campaign (HDF--S; Williams et~al.
2000) is distinguished in part from the hugely successful Hubble Deep
Field-North (HDF--N; Williams et~al.\ 1996) by ultraviolet (UV)
spectroscopy (Ferguson et~al.\ 2000) of the bright, radio-quiet quasar
J2233--6033 (HDF--S QSO; $V \sim 17.5$, $z = 2.24$; Boyle 1997), using
the Space Telescope Imaging Spectrograph (STIS; Woodgate et~al.\
1998). Medium- and high-resolution STIS spectroscopy provides detailed
coverage of the Ly$\alpha$ forest from $1.2 < z <1.93$. Optical
spectroscopy from the ground (Outram et~al.\ 1999, Sealey et~al.\
1998, Savaglio et~al.\ 1998, Cristiani et~al. 2000) extends coverage
of the Ly$\alpha$ forest to $z =$ 2.24 and also covers various metal
lines along the QSO line of sight. The UV spectrum of the HDF--S QSO
provides a unique window on the universe over this redshift interval
where very little is known due to observational difficulties.

The HDF--S QSO spectra have revealed several features in the
distribution of Ly$\alpha$ gas clouds (Savaglio et~al.\ 1999),
including a ``void'' spanning $z=1.383-1.460$ (94 co-moving ${\rm
h}_{65}^{-1}$ Mpc)\footnote{We assume \Ho\ = 65 ${\rm
km~s^{-1}~Mpc^{-1}}$, \qo\ = 0.5, $\Lambda=0$, and co-moving distances
throughout this paper.}, and two overdensities at $z\sim1.335$ and
$z=1.92-1.99$ (covering 64 ${\rm h}_{65}^{-1}$ Mpc).  Both Ly$\alpha$
overdensities have associated metal systems: one tentatively
identified at $z=1.336$ (Ferguson et~al.\ 2000), and two at $z=1.929$
and $z=1.943$ (Savaglio et~al.\ 1998).  There is a QSO, J2233--6032,
at a redshift $z=1.335$ and a projected distance of 44\arcsec\ (0.7
${\rm h}_{65}^{-1}$ Mpc) from the HDF--S QSO (Tresse et~al.\ 1998). In
addition, there are three galaxies with redshifts of $z =$ 1.268,
1.315 and 1.218 at a distance $\lesssim 2 {\rm h}_{65}^{-1}$ Mpc from
the HDF--S QSO line-of-sight (Glazebrook et~al.\ 2000).  A
concentrated ground based search for other objects associated with
these features will be required to better understand their nature.

We have performed a multi-color ($uBVRI$+narrow-band) imaging survey
of 1/2 square degree centered on the HDF--S (Palunas et~al.\ 2000).
The narrow-band filter (NB) was centered at 3958\AA\ to detect
Ly$\alpha$ at the redshift of the HDF--S QSO.  Our goals for the
survey include determining photometric redshifts of galaxies over a wide
field around the HDF--S , searching for emission-line objects near
the HDF--S QSO, and identifying other quasars in the field.  Here we
present candidate quasars in our catalog. We confirm the presence of
Ly${\alpha}$ nebulosity extending $\sim12\arcsec$ around the HDF--S
QSO, reported by Bergeron et~al. (1999). We detect 10 point-like
objects in emission through the NB filter. Of these, 7 satisfy our QSO
color selection criteria.  We report the discovery of a $B\sim20$
quasar at $z=1.56$ just 6.7\arcmin\ (6.8 ${\rm h}_{65}^{-1}$ Mpc) from
the HDF--S QSO line of sight and $\sim$12\arcsec\ from the western
edge of the WFPC2 field. Images and catalogs from the survey are
available on the world wide web at
http://hires.gsfc.nasa.gov/$\sim$gardner/hdfs.

\section{Observations and Reductions}

We observed the HDF--S region using the Big Throughput Camera (BTC) on
the CTIO 4m during 1998 September.  A complete description of the
reductions can be found in Palunas et~al.\ (2000).  Images were taken
in the Sloan Digital Sky Survey $u$, Johnson $B$~and $V$, and Cousins
$R$~and $I$~filters. We also imaged the field using a narrow-band
interference filter centered at 3958\AA\ and 50 \AA\ wide tuned to
Ly$\alpha$ at the redshift of the HDF--S QSO. The BTC is a prime-focus
mosaic camera consisting of 4 CCDs, each 2048$\times$2048 pixels
(Wittman et~al.\ 1998). The camera has a 0.43\arcsec\ pixel size and
covers an overall area of (34.8\arcmin)$^2$ with 5.4\arcmin\ gaps
between the CCDs. Individual exposures were dithered to fill in the
gaps between the CCDs, which resulted in a contiguous field with
non-uniform coverage. The field extends $44.0\arcmin \times 44.0
\arcmin$; the total area covered, excluding regions around bright
stars, is 1716 $\Box^{\arcmin}$. The NB imaging covers less area,
1352$\Box^{\arcmin}$. 

The conditions were photometric.  The photometry was calibrated using
Landolt standards, and fixed to the Johnson $UBV$~and Cousins
$RI$~photometric system as defined by Landolt (1992).  Comparison of
the Johnson $U$~and Sloan $u$~filters demonstrates that small color
corrections are sufficient to place our observations onto the same
system as Landolt.

The seeing was 1.2\arcsec--1.5\arcsec. Bias, dark subtraction and
flatfielding of the images using superflats, were done using
IRAF\footnote{IRAF is distributed by NOAO, which is operated by AURA
Inc., under contract to the NSF.}. Astrometric corrections for the
wide-field optical distortions were derived and the corrected images
were combined using \emph{btcproc} (Wittman et~al.\ 1998). Catalogs of
individual sources were compiled using SExtractor (Bertin \& Arnouts
1996). The limiting magnitudes vary across the images.  The 5$\sigma$
detection limits for point-like objects in each color range between
$24<u<25$, $25.6<B<26.6$, $25<V<26$, $25<R<25.8$, $23.5<I<24.4$, with
about 50\% of the area of each image at the brightest limits.  For the
NB image the 5$\sigma$ limits range between 0.5---1.2 $\times
10^{-16}~{\rm erg~s^{-1}~cm^{-2}}$.

\section{Quasar candidates}
Most quasars are easily distinguished from other point sources on the
basis of their unusual colors, particularly in the UV (e.g. Hall
et~al.\ 1996a, 1996b).  We transform our colors from the Landolt to
Cousins photometric system that Hall et~al. use (Bessel 1991), and
denote the transformed colors with a subscript C.  The corrections are
small and do not affect the results.  Figure \ref{fig_B-V_u-B} shows
the $(u-B)_{\rm C}$ vs. $(B-V)_{\rm C}$ colors of the point sources in
our catalog.  Our QSO selection criteria are given in Table 1.  Hall
et~al.'s selection criteria are designed to exclude the dense stellar
locus running diagonally across the diagram and compact emission-line
galaxies which populate the upper right of the diagram. Their criteria
are adjusted as a function of magnitude to account for broadening of
the stellar locus due to photometric errors. Our criteria differ
slightly at fainter magnitudes because our photometry is deeper and
therefore the stellar locus exhibits less broadening.

Point sources in our data were selected on the basis of the FWHM and
``class star'' (cl) parameters from SExtractor. The selection criteria
were determined by comparing to the sources classified in the HDF--S
flanking fields (Lucas, R. et~al. 2000). The following parameters
maximized the number of point sources selected without excessive
contamination by extended sources: for $R \leq 20$, FWHM $<
1.8$\arcsec; for $20 < R \leq 23$, FWHM $<$ 1.8\arcsec\ and cl $>
0.95$; and for $R > 23$, FWHM $< 1.9\arcsec$ and cl $> .92$.  Full
discussion will be provided elsewhere (Palunas et~al. 2000, in
preparation).

Using the above criteria we find 4 $B < 19$, 13 $B < 20$, 30 $B < 21$,
71 $B < 22$ and 154 $B < 23$ QSO candidates in our field.  Figure
\ref{fig_xy} shows the distribution of QSO candidates in the
field. The QSO selection criteria identify quasars at $z<2.3$. Colors
for QSOs at higher redshifts lie within the stellar locus.  Hall
et~al.\ find a 60\% QSO selection efficiency for objects with these
colors at ${B < 22}$. In Figure \ref{fig_numb} we compare the
integrated number counts of the QSO candidates we have selected,
assuming 60\% selection efficiency, to the QSO counts of Hartwick \&
Shade (1990).

\section{The HDF--S QSO}
Bergeron et~al. (1999) report large scale extended Ly$\alpha$
nebulosity around the HDF--S QSO, detected in narrow-band images taken
at the VLT. The nebulosity is reported to extend approximately
9.2\arcsec\ $\times$ 12.1\arcsec; the outer parts are patchy but
distributed symmetrically around the QSO. They find no evidence of a
continuum component associated with the nebulosity. Emission-line
nebulae of this size and luminosity are common around radio-loud
quasars, but not around radio-quiet quasars such as the HDF-S
QSO. However, the nebulae around radio-loud QSOs are generally
asymmetric and roughly aligned with the radio jets.

We confirm the existence of extended Ly${\alpha}$ emission around the
HDF-S QSO. Figure \ref{fig_neb} shows the narrow-band luminosity
profile of the HDF--S QSO compared with the profiles of 8 stars within
$\sim$3\arcmin\ of the QSO.  The stellar profiles are normalized to
show the same flux as the QSO within a 4\arcsec\ radius aperture. The
images suffer from low-level ghost reflections off the NB filter; the
manufacturer dropped a completed filter shortly before our observing
run and the replacement did not have an AR coating. The ghosts are
well modeled by a Gaussian which we subtract from the profiles. The
peak amplitude of the ghost is $1.75\times10^{-4}$ times the stellar
flux within a 4\arcsec\ radius and its width is
$\sigma=3.6\arcsec$. The flux from the ghost peaks at a radius of
$\sim 9.5$arcsec.  The QSO profile exhibits significant excess flux
for $r>4\arcsec$ when compared to all 8 stellar profiles even where
the flux from the model ghost is negligible.

The integrated Ly$\alpha$ flux in an annulus with
3\arcsec$<r<$10\arcsec\, centered on the QSO is $(2.9\pm0.5)\times
10^{-15}~{\rm erg~s^{-1}~cm^{-2}}$. Because the low flux levels and
wide area make absolute flux measurement difficult we estimate the
error by scaling the 1$\sigma$ deviation of the normalized relative
flux of the 8 stars integrated over the same area as the QSO.
Bergeron et~al. (1999) report a flux of 3.2$\times10^{-15}~{\rm
erg~s^{-1}~cm^{-2}}$.

\section{Emission-line objects}

To identify point-like emission-line objects in the field we compare
the NB magnitudes to an estimate of the continuum around the NB
filter.  We estimate the continuum as the average $u$- and $B$-band
flux, $\langle uB\rangle$. In Fig. \ref{fig_colmag} we plot the
$\langle uB\rangle-$ NB color-magnitude diagram for the point sources.
The wide spread in $\langle uB\rangle-$ NB colors is due to the
4000\AA\ break in spectra of the stars.

To set limits for emission-line detection we have estimated the upper
99.5\% confidence limit on the $\langle uB\rangle$ - NB~colors of
non-detections using Monte Carlo simulations (Teplitz et~al.\ 1998). A
Gaussian distribution of errors is applied to the photometry for
simulated line-free objects, and the confidence intervals are measured
as a function of broad-band magnitude. We used the global average of
the errors over the entire frame for the broad and narrow band data.

We detect 10 spatially unresolved, emission-line sources in our NB
filter including the HDF--S QSO.  Of these, 7 are QSO candidates by
our criteria. Table 2 lists the emission-line objects. Figure
\ref{fig_xy} shows their position in the field. The observed
equivalent widths are lower limits since part of the broad-line
emission may fall outside of the narrow-band filter.  Figure
\ref{fig_sed} plots the spectral energy distributions of the objects
in AB magnitudes.  We include JHK photometry (da Costa et~al.\ 1998)
for objects for which it is available.

\subsection{QSO B}

In addition to the emission we detect at 3958\AA, Glazebrook et~al.\
(2000) detect a broad emission line at 7174\AA\ for QSO candidate B
(see Table 2).  The object is unresolved in an HDF--S flanking-field
image.  The emission lines are identified as $[$\ion{C}{4}$]$
$\lambda$1549 at $z=1.56\pm0.02$ and \ion{Mg}{2} $\lambda$2800 at
$z=1.56\pm0.01$. We adopt a redshift of $1.56\pm0.01$.  Relatively
small systematic biases in the redshifts determined from these lines
(e.g. Tytler \& Fan 1992, Laor et~al. 1995) cannot be resolved.  The
absolute $B$-band magnitude is ${\rm M}(B) = -25$ using the Veron \&
Veron (1998) prescription.  The rest-frame equivalent width of the
\ion{C}{4} emission doublet is at least 60.5 \AA. The rest-frame
equivalent width of the \ion{Mg}{2} lines is 47 \AA.  The FWHM of the
\ion{Mg}{2} line is 82 \AA\ in the rest frame, compared to an
instrumental resolution of 3 \AA\, establishing it as a broad line
typical of QSOs.  There is an excess H-band flux which is probably due
to H$\alpha$ emission.  The equivalent width required to give the
observed excess above the J and K bands is $\sim$240 \AA\ in the rest
frame. This is near the minimum observed equivalent width in quasars
(Espey et~al.\ 1989). The object is detected at 7$\mu$m (5$\sigma$)
and 14$\mu$m bands (3$\sigma$) with the Infrared Space Observatory
(ISO) (Oliver et~al.\ 2000), but flux calibrated magnitudes are not
yet available.  The QSO is radio quiet; it is detected with a flux of
163 $\mu$Jy at 4.9 GHz and 111 $\mu$Jy at 8.6 GHz in observations at
the Australia Telescope National Facility (ATNF) HDF--S (Norris
et~al.\ 2000).

QSO B is 6.7\arcmin\ (6.8 ${\rm h}_{65}^{-1}$ Mpc) from the line of
sight to the HDF--S QSO and $\sim$12\arcsec\ from the western edge of
the WFPC2 field. There is no statistically significant excess of
absorption lines in the HDF--S QSO associated with QSO B.  Since it is
so close to the HDF--S, this quasar provides an important new
line-of-sight to study the $z<1.56$ features revealed in the HDF--S
HST images and QSO spectrum.

\subsection{Other Objects}

QSO emission lines which could be detected in our NB images are
Ly$\alpha$ at $z\sim2.26$, \ion{C}{4} $\lambda$1549 at $z\sim1.55$,
\ion{C}{3} $\lambda$1909 at $z\sim1.07$ or \ion{Mg}{2} $\lambda$2791
at $z\sim0.42$.  The NB images sample 6.24$\times$, 5.10$\times$,
3.69$\times$ and 1.24$\times10^{4}~{\rm h}_{65}^{-3}{\rm Mpc}^3$ at
each of these redshifts, respectively.  The effective volume sampled
could be as much as 50\% higher depending on how broad the emission
lines are.  Compact emission-line galaxies could be detected in
\ion{O}{2} at $z\sim0.06$. To estimate the space density of QSOs we
integrate the (Hawkins \& Veron 1995) luminosity function, with a
constant logarithmic slope of 0.63, down to M$_B=-23$ which is the
border between objects classified as QSOs or Seyfert 1 nuclei (Schmidt
\& Green 1983) and about our detection limit for QSOs at $z=2.2$.  The
expected space density of QSOs in units of ${\rm h}_{65}^{3} {\rm
Mpc}^{-3}$ is 2.4$\times10^{-5}$ at $2.2<z<3.2$, 1.5$\times10^{-5}$ at
$1.2<z<1.7$ and 0.5$\times10^{-5}$ at $0.7<z<1.2$ in our cosmology.
At $z < 1.5$, the amplitude of the luminosity function falls as $(1+
{\rm z})^{-6}$. Under this assumption the expected space density at
redshift z$=0.4$ is roughly 30 times less than that at z$=1.5$ or
$\sim 5\times10^{-7}~{\rm h}_{65}^{3} {\rm Mpc}^{-3}$.  The number of
QSOs expected in our detection windows is about 2.5: 1.5 in
Ly$\alpha$, 0.8 in \ion{C}{4} and 0.2 in \ion{C}{3} and 0.005 in
\ion{Mg}{2}. We detect 7 emission-line objects which we identify as
QSO candidates excluding the HDF--S QSO.  The total number of
detections agrees with the expected number to within the 3$\sigma$
Poisson errors (Gehrels 1986).  Due to small number statistics there
is no evidence of an overdensity in any of the redshift windows.

\section{Summary}

We have presented a sample of QSO candidates in the HDF--S region,
including one confirmed QSO, at redshift $z=1.56$ just 6.7\arcmin\
(6.8 ${\rm h}_{65}^{-1}$ Mpc) from the HDF--S QSO. We confirm the
existence of a large, 24\arcsec\ diameter (490~${\rm h}_{65}^{-1}$
kpc) nebular emission surrounding the HDF--S QSO. The existence of
such a cloud around a radio-quiet QSO is highly unusual. QSOs in the
HDF--S region will provide an important tool to map structure over a
much larger area than that covered by the HST data.  Absorption line
spectra of the brighter QSOs will allow detection of low column
density gas clouds along lines of sight other than the HDF--S QSO.

\acknowledgments

We thank Dave Wittman and Steve Kraemer for useful discussions.  We
warmly acknowledge the excellent support of the CTIO observing staff.
This work was funded by the National Research Council, by NASA grant
NRA--98--03--UVG--011, and by NASA funding to the STIS Instrument
Definition Team (459--10--60).  Finally, we thank the anonymous
referee for a rapid response and very helpful comments.

\clearpage

\begin{deluxetable}{rl}
\tablenum{1}

\tablecolumns{2}
 
\tablecaption{QSO Color Selection}
\tablehead{
\colhead{Mag}&
\colhead{Color Limits}
}

\startdata

$B <21$          &  $(B-V)_{\rm C} < 0.35$\phn~for $(u-B)_{\rm C}\geq -0.3$ and    \\
                 &  $(B-V)_{\rm C} < 0.6$\phn\phn~for $(u-B)_{\rm C} < -0.3$          \\
$21\leq B <22$   &  $(B-V)_{\rm C} < 0.325$~for $(u-B)_{\rm C} \geq -0.35$ and  \\
                 &  $(B-V)_{\rm C} < 0.6$\phn\phn~for $(u-B)_{\rm C} < -0.35$         \\
$22\leq B <23$   &  $(B-V)_{\rm C} < 0.3$\phn\phn~for $(u-B)_{\rm C} \geq -0.4$ and   \\
                 &  $(B-V)_{\rm C} < 0.5$\phn\phn~for $(u-B)_{\rm C} < -0.4$          \\
\enddata

\end{deluxetable}

\clearpage

\begin{deluxetable}{cllllrl}
\tablenum{2}

\tablecolumns{7}
 
\tablecaption{Narrow-Band Detections}
\tablehead{
\colhead{ID}&
\colhead{R.A. (J2000)}&
\colhead{Dec. (J2000)}&
\colhead{$B$}&               %     \tablenotemark{1}}&
\colhead{QSO}&
\colhead{E.W. (\AA)}&
\colhead{notes} 
}

\startdata

HDF-S & 22:33:37.59  &  -60:33:29.06 & 17.39  & y   &  119.8 & z=2.24\\
A     & 22:35:07.39  &  -60:37:18.72 & 18.47  & y   &   78.9 & ...\\
B     & 22:32:43.47  &  -60:33:51.55 & 20.31  & y   &  120.8 & z=1.56\\
C     & 22:35:10.47  &  -60:35:30.33 & 20.51  & y   &   73.4 & ...\\
D     & 22:36:09.84  &  -60:43:36.12 & 21.22  & y   &   78.0 & ...\\
E     & 22:36:07.77  &  -60:33:20.40 & 21.51  & n   &  100.3 & ...\\
F     & 22:32:44.83  &  -60:48:16.38 & 21.59  & y   &  134.9 & ...\\
G     & 22:32:32.78  &  -60:18:43.14 & 21.76  & y   &  107.0 & ...\\
H     & 22:31:14.08  &  -60:18:25.25 & 22.07  & n   &  100.8 & ...\\
I     & 22:33:00.37  &  -60:19:11.46 & 22.85  & y   &  118.8 & ...\\
J     & 22:32:14.83  &  -60:54:30.23 & 23.74  & R.G. &  227.8 & ...

\enddata

\end{deluxetable}

\clearpage

%[Palunas.fig1.ps]
\figcaption{$(u-B)_{\rm C}$ vs. $(B-V)_{\rm C}$ diagram for point
sources with $B<23$.  For clarity the error bars are drawn on
one side only.  The confirmed QSOs, the HDF--S QSO, QSO B and the
Tresse et~al. QSO are marked with filled circles. The rest of the
emission-line candidates are marked with an open circles. Objects
detected in radio (Norris et~al.\ 2000) are marked with stars. The QSO
selection criteria are discussed in the text.  {\label{fig_B-V_u-B}}}

%[Palunas.fig2.ps]
\figcaption{The distribution of quasar candidates in our field.  The
size of the markers are proportional to the magnitude.  Candidates
identified with radio sources (Norris et~al.\ 2000) are marked with
stars.  Objects identified as emission-line sources are marked with a
large open circle and labeled (see Table 2). The image shows the
HDF--S fields for STIS, WFPC2 and NICMOS (solid lines), as well as the
WFPC2 flanking fields (dotted line). Parts of the image excluded from
analysis due to bright stars are shaded. {\label{fig_xy}}}

%[Palunas.fig3.ps]
\figcaption{QSO integrated number counts. The solid line is the number
counts for our $(u-B)_{\rm C}$, $(B-V)_{\rm C} $ selected QSO
candidates assuming a 60\% selection efficiency and excluding the
HDF--S QSO. The open triangles are the number counts of Hartwick
\& Shade (1990) for QSOs at $z<2.2$. {\label{fig_numb}}}

%[Palunas.fig4.ps]
\figcaption{Luminosity profile for the HDF--S QSO is compared to the
mean normalized PSFs of 8 nearby stars. The shaded region around the
mean PSF profile shows the 1$\sigma$ scatter of the 8 PSFs.  A model of
a ghost reflection was subtracted from all the profiles. The ghost
amplitude was scaled to be a fixed ratio of the flux within a
4\arcsec\ radius aperture of each object. {\label{fig_neb}}}

%[Palunas.fig5.ps]
\figcaption{Narrow--Continuum color magnitude diagram. The continuum
is estimated by the average of the $u$ and $B$ flux, $\langle
uB\rangle$. The line represents the upper 99.5\% confidence limit on
the $\langle uB\rangle$ - NB~colors of non-detections. Candidate QSOs
selected in this paper (see Fig.\ref{fig_B-V_u-B}) are marked with
small solid circles and radio sources (matched to Norris et~al.\ 2000)
are marked with stars.{\label{fig_colmag}}}

%[Palunas.fig6.ps]
\figcaption {The SEDs of all the NB emission-line objects. Our
broad-band and NB data is traced with a solid line. The dashed line
traces the optical and near IR data of da Costa et~al.\ (1998).  The
thin lines are the spectra of Sealey et~al.\ 1998 for the HDF--S QSO
and Glazebrook et~al.\ (2000) for QSO B shifted below the SEDs for
clarity.  The QSO candidates are labeled with Q.{\label{fig_sed}}}

\clearpage
\begin{figure}[h]
\plotone{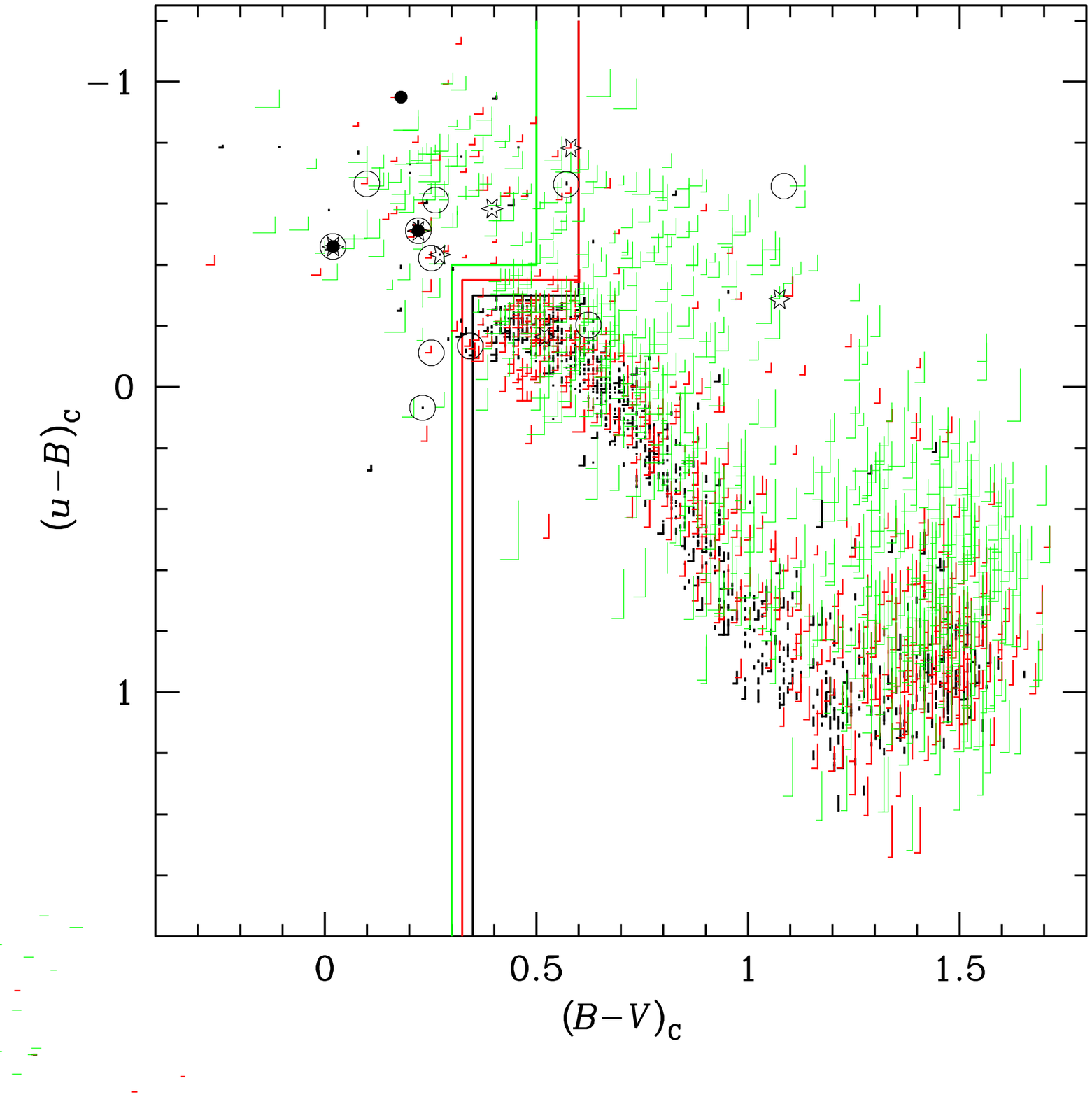}
\end{figure}

\clearpage
\begin{figure}[h]
\plotone{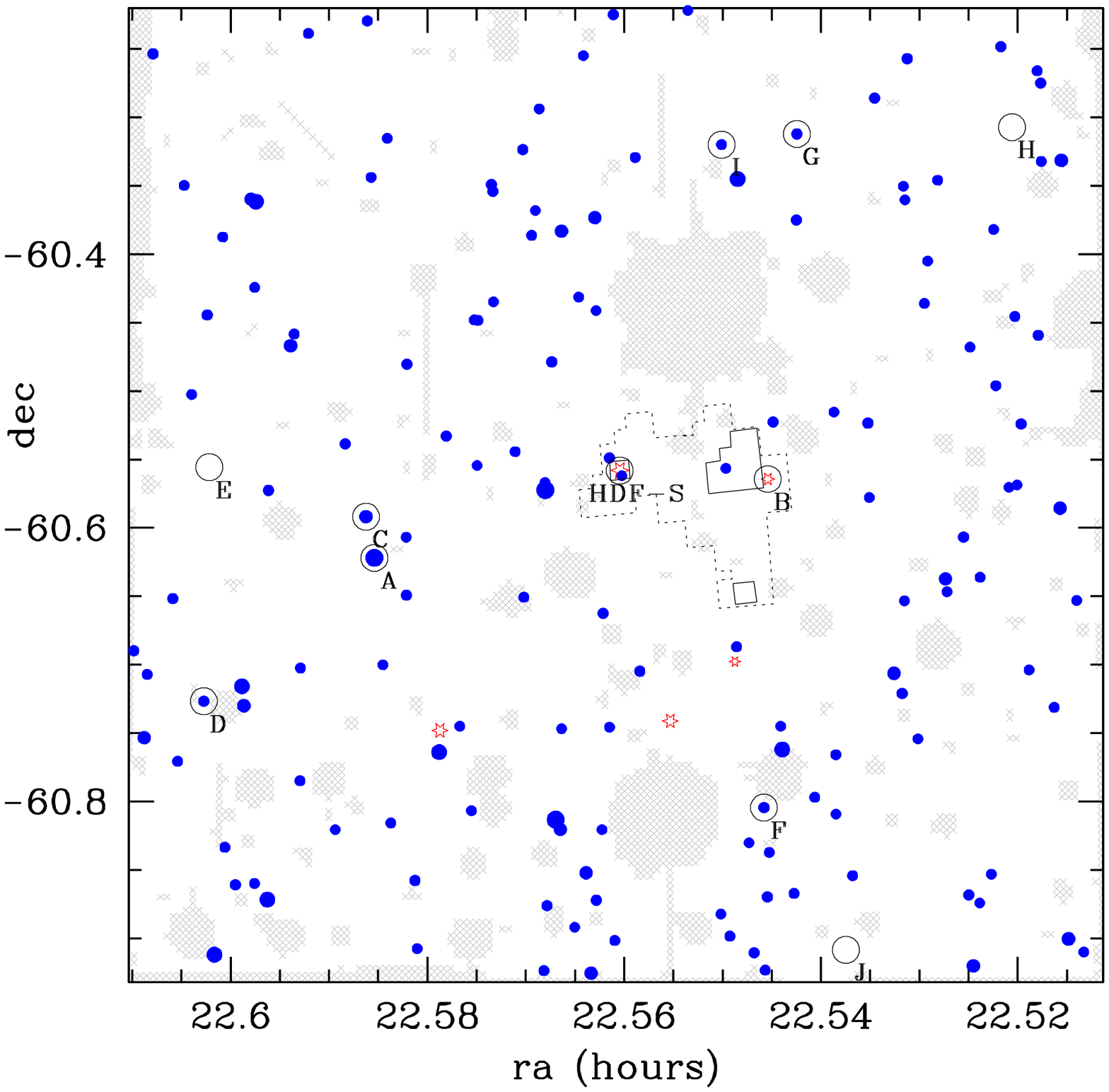}
\end{figure}

\clearpage
\begin{figure}[h]
\plotone{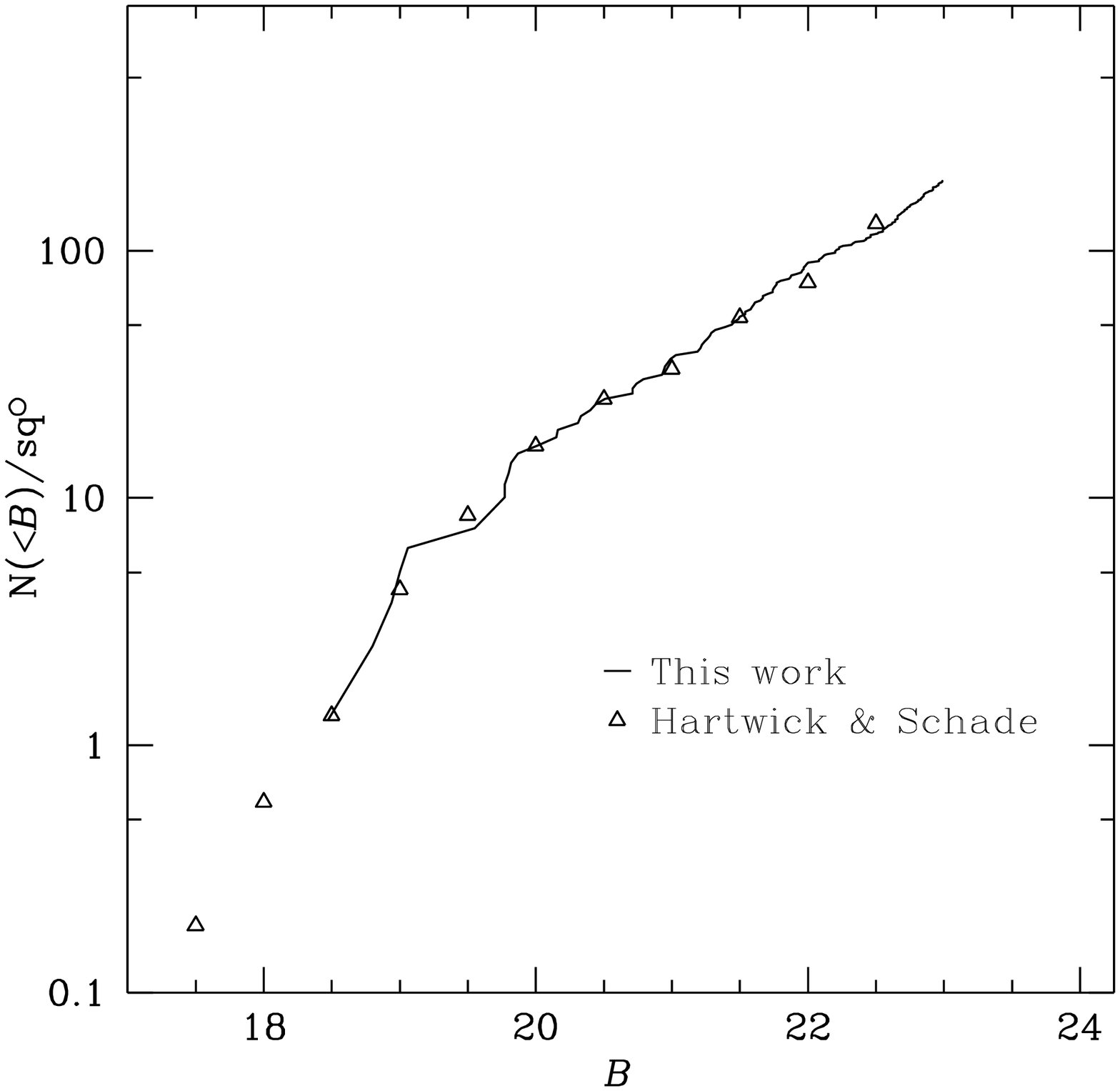}
\end{figure}

\clearpage
\begin{figure}[h]
\plotone{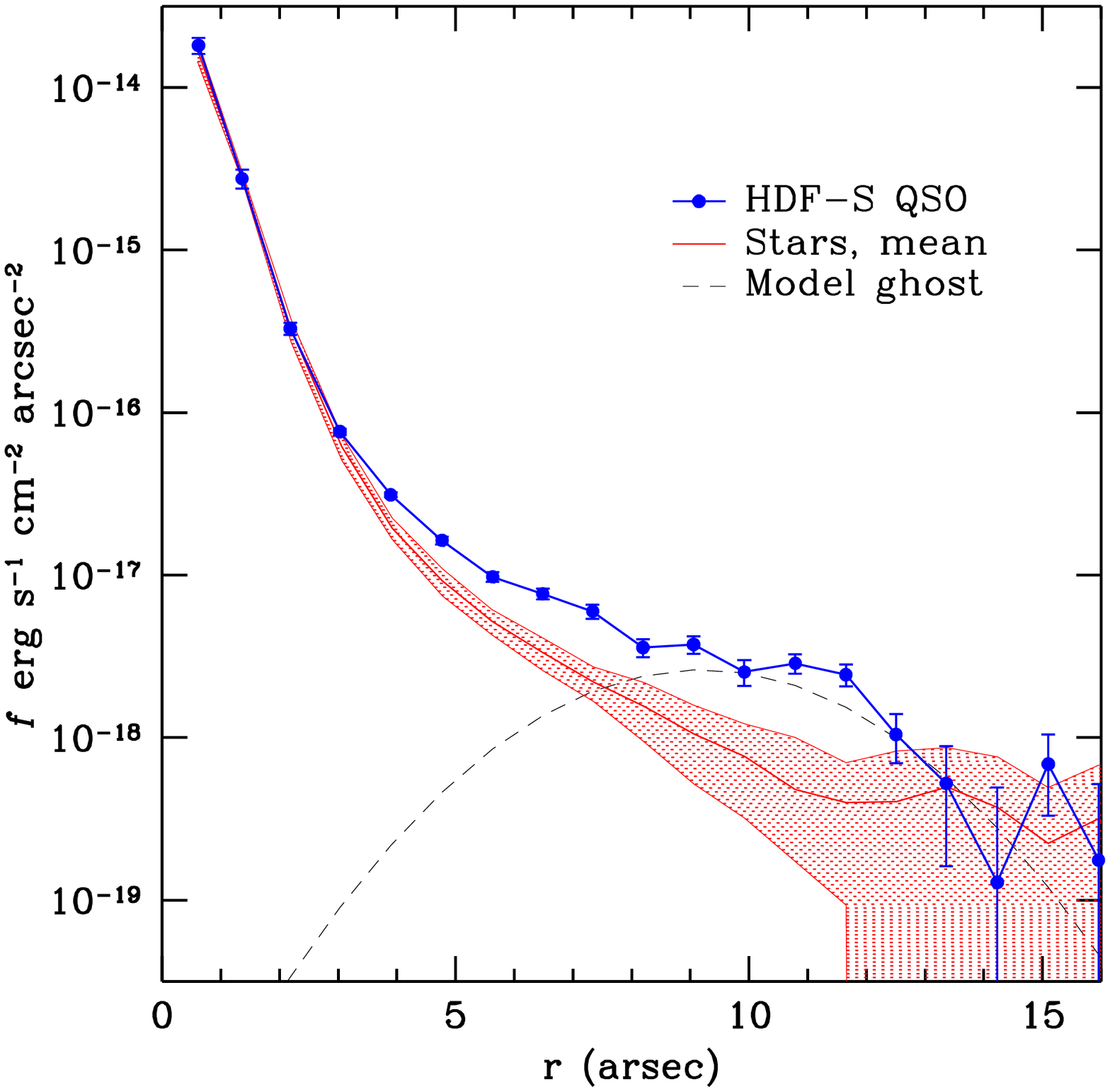}
\end{figure}

\clearpage
\begin{figure}[h]
\plotone{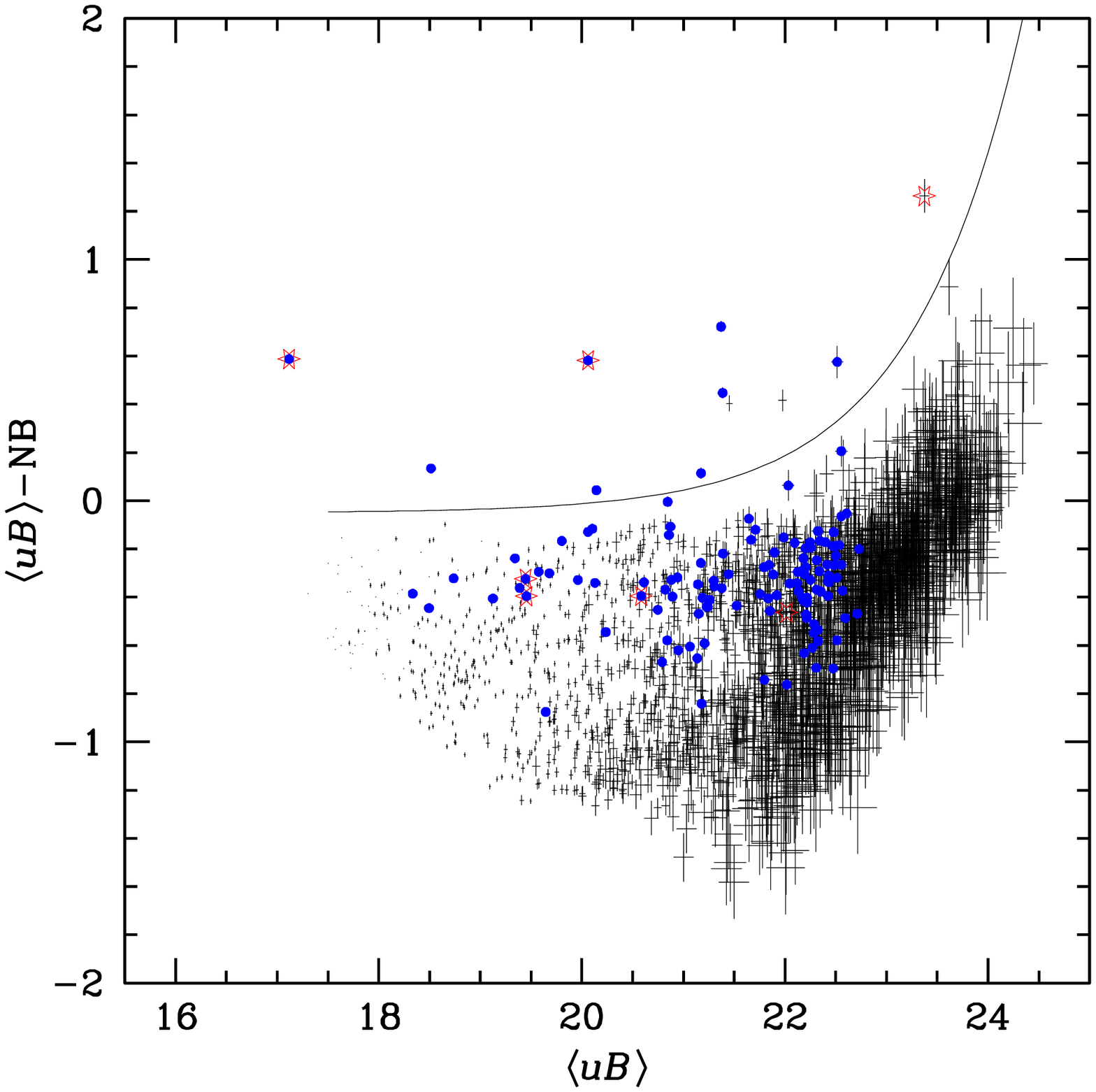}
\end{figure}

\clearpage
\begin{figure}[h]
\plotone{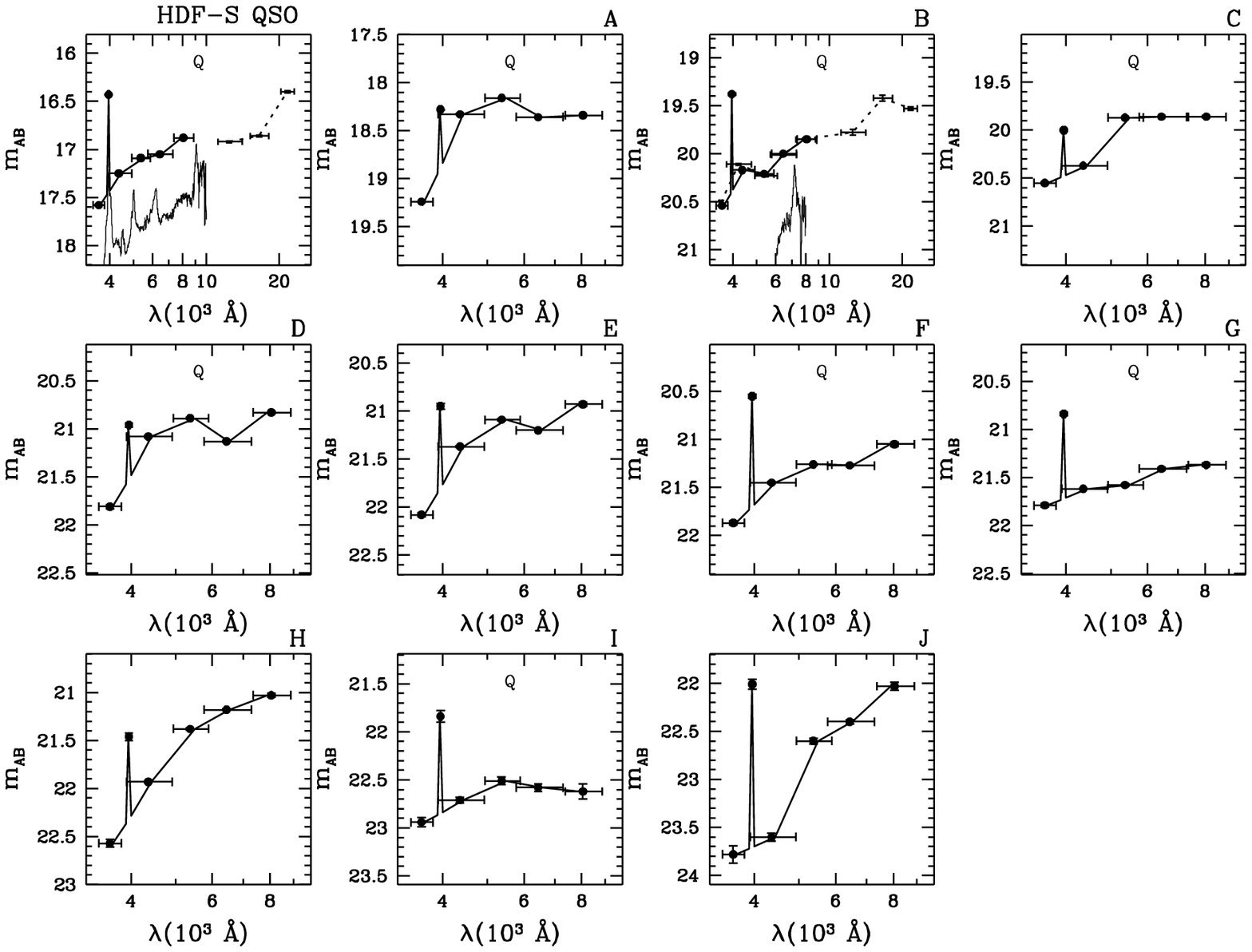}
\end{figure}

\end{document}